\documentclass[]{JHEP}

\usepackage{epsfig}

\def\a{& \hspace{-7pt}}
\def\c{\hspace{-5pt}}
\def\bea{\begin{eqnarray}}
\def\eea{\end{eqnarray}}
\def\be{\begin{equation}}
\def\ee{\end{equation}}
\def\nn{\nonumber}

\def\Z{{\bf Z}}

\newcommand{\eqalign}[1]{\hspace{-10pt}\begin{array}{ll}#1\end{array}
\hspace{-10pt}}
\title{Gauge and gravitational anomalies in $\mathbf{D=4}$ 
$\mathbf{N=1}$ orientifolds}
\author{Claudio A. Scrucca \\
Sektion Physik, Ludwig Maximilian Universit\"at M\"unchen\\
Theresienstra\ss e 37, 80333 Munich, Germany\\
{\footnotesize \tt Claudio.Scrucca@physik.uni-muenchen.de}}
\author{Marco Serone \\
Department of Mathematics, University of Amsterdam \\
Plantage Muidergracht 24, 1018 TV Amsterdam, The Netherlands; \\
Spinoza Institute, University of Utrecht \\
Leuvenlaan 4, 3584 CE Utrecht, The Netherlands \\
{\footnotesize \tt serone@wins.uva.nl}}
\abstract{The cancellation of $U(1)$-gauge and $U(1)$-gravitational 
anomalies in certain $D=4$ $N=1$ Type IIB orientifolds is
analyzed in detail, from a string theory point of view. 
We verify the proposal that these anomalies are cancelled by a 
Green-Schwarz mechanism involving only twisted Ramond-Ramond fields.
By factorizing one-loop partition functions, we also get the
anomalous couplings of D-branes, O-planes and orbifold fixed-points 
to these twisted fields. Twisted sectors with fixed-planes participate 
to the inflow mechanism in a peculiar way.}

\preprint{UvA-WINS-Wisk-99-20 \\ SPIN-1999/30 \\ LMU-TPW 99-21\\ 
{\tt hep-th/9912108}}
  
\begin{document}


\section{Introduction}

Thanks to the discovery of D-branes and the understanding of their 
prominent role in string theory \cite{pol}, new interesting possibilities 
have opened up to construct phenomenologically viable string vacua. 
Among others, four dimensional $N=1$ Type IIB orientifolds 
\cite{abpss,ks1,ks2,afiv,z,kst} represent an example of 
new perturbative models that have become accessible in alternative
to the well-studied heterotic compactifications.
An interesting issue in these new vacua is anomaly cancellation.
It has been proposed and argued in \cite{iru1} that both $U(1)$-gauge 
and $U(1)$-gravitational anomalies are cancelled by a generalization of 
the $D=4$ Green-Schwarz (GS) mechanism \cite{dsw}, involving the exchange 
of twisted Ramond-Ramond (RR) closed string states only.
This proposal has been analyzed in detail at the level of low-energy 
effective action in \cite{klein} (see also \cite{celw,cpw,lln}).
A string theory analysis of some of the anomaly cancelling terms 
for these orientifolds has been given in \cite{abd}, which is 
however limited to special models and for gauge anomalies only.
Due to the potential relevance of these string vacua in building 
realistic models (see for instance \cite{aiq}), it is of 
interest to perform a more complete analysis of the cancellation
of anomalies in these models at the string level.

In \cite{ss2}, a general method for the study of anomalies 
in string theory vacua, based on the computation of the topological 
one-loop partition function in the presence of 
gauge and gravitational backgrounds, has been described.
At lowest order in derivatives (momenta), this calculation yields 
directly the one-loop anomaly from the charged massless (open string) 
spectrum and {\it at the same time} (minus) the tree-level inflow of 
anomaly mediated by neutral massless (closed string) fields. 
By analyzing the transverse channel, it is possible to determine 
which states participate to the GS mechanism. On the other hand, the direct 
channel analysis allows a precise check of the charged spectrum.

Our results are in agreement with the proposal of \cite{iru1}: 
the only fields participating to the inflow of anomaly are combinations
of twisted RR axions (with twists different from ${\bf Z}_2$).
Notice that this is in striking contrast with what happens
in six-dimensional $N=1$ IIB orientifolds, where it was explicitly shown
in \cite{ss2} that all closed RR string states, both twisted and untwisted,
participate in general to the anomaly cancellation mechanism. 
We also derive by factorization of the above one-loop partition functions,
the Wess-Zumino (WZ) couplings for D9-branes, D5-branes and orbifold
fixed-points to the twisted RR axions. Interestingly, the gravitational 
couplings for fixed-points can be completely reabsorbed in those of the 
D-branes present in the models, with the net effect of rescaling by a factor 
of $3/2$ the gravitational part. 

For the $N=1$ sectors, {\em i.e.} those without planes left fixed by 
the orbifold action, all the anomalous couplings can then be rewritten 
in a unified way, as reported in the formulae (\ref{GSN}) and (\ref{GSNM}). 
The $N=2$ sectors, containing fixed-planes, are 
instead more subtle. In these cases, a non-vanishing 
inflow of anomaly arises only in the 95 sector and correspondingly
it is not possible to fix unambiguously the anomalous couplings
by factorization. However, one can conclude that neither the D9-branes nor 
the D5-branes can couple to the simple and natural symmetric combination of 
the corresponding RR twisted axions, as happens in $N=1$ sectors.
Although we do not have a satisfactory and precise explanation of this 
fact, we will see that the general form of these couplings might allow
interesting tree-level corrections to the gauge couplings, even for unbroken 
non-Abelian gauge group. This is in contrast to the situation in the 
$N=1$ sectors, where the presence of Fayet-Iliopoulos terms, related by 
supersymmetry to some of the WZ terms above, fixes to zero the tree-level
gauge couplings dependence on the twisted Neveu-Schwarz-Neveu-Schwarz (NSNS)
scalars, supersymmetric partners of the RR axions.
Without entering into all the details of the low-energy effective action, 
which has been extensively analyzed in \cite{iru1,abd,klein}, we also
discuss the spontaneous breaking of $U(1)$ factors through a Higgs mechanism 
induced by these anomaly-cancelling couplings, both in the $N=1$ and 
$N=2$ sectors.

The paper is organized as follows. In section two, we briefly review
some properties of the orientifold models under analysis. In section three, 
we compute the one-loop partition function in the odd spin-structure
yielding the inflow of anomaly. In section four, we deduce then the WZ 
couplings by factorization. The last section contains a brief field theory 
analysis of our results. Finally, we report in the appendix the 
explicit combinations of $U(1)$ gauge fields which become massive.


\section{$D=4$ $N=1$ orientifolds}

In this section, we review some generalities about the $D=4$ $N=1$ 
Type IIB orientifolds we consider.
These models always contain 32 D9-branes, required to cancel the tadpoles
from the O9-plane associated to the world-sheet parity operator $\Omega$, 
and 32 D5-brane wrapped along the third compact plane when $N$ is even, 
required to cancel the tadpoles from the 16 O5-planes associated to the 
element $\Omega R$ of the orientifold group, $R$ being a reflection along 
the first two compact planes. 
In the following, we shall restrict to the maximally symmetric case in 
which all the D5-branes sit at the origin of the first two compact planes.

\vskip 10pt 
\noindent 
{\bf $\Z_N$ orientifolds}
\vskip 5pt 

\noindent
The $\Z_N$ action is generated by the element $\theta = exp (2 \pi i v_i J_i)$,
where $J_i$ is the rotation generator in the $i$-th compact plane and
$v_i$ are the corresponding components of the twist vector defining the 
action, $v = (v_1,v_2,v_3)$. In the open string sector, the twist 
$\theta^k$ is represented by matrices $\gamma_k$ on the Chan-Paton wave 
function. In a suitable basis, one can choose 
$\gamma_{k,9} = \gamma_{k,5} = (\gamma)^k$.
The $\Z_N$ actions leading to consistent models with cancelled tadpoles 
are given in the table below.

\begin{table}[h]
\vbox{
$$\vbox{\offinterlineskip
\hrule height 1.1pt
\halign{&\vrule width 1.1pt#
&\strut\quad#\hfil\quad&
\vrule#
&\strut\quad#\hfil\quad&
\vrule#
&\strut\quad#\hfil\quad&
\vrule width 1.1pt#\cr
height3pt
&\omit&
&\omit&
&\omit&
\cr
&\hfil $\c G \c$&
&\hfil $v$&
&\hfil $\gamma$&
\cr
height3pt
&\omit&
&\omit&
&\omit&
\cr
\noalign{\hrule height 1.1pt}
height3pt
&\omit&
&\omit&
&\omit&
\cr
&\hfil $\c\Z_3\c$&
&$\eqalign{(1,1,-2)/3}$&
&$\eqalign{
{\rm diag}\left(\alpha^{2}{\bf I_{12}\raisebox{9pt}{}}^{\!\!\!a}, 
\alpha^{- 2}{\bf I_{\,\overline{\!12\!}\,}\raisebox{9pt}{}}^{\!\!\!a},
{\bf I_{8}\raisebox{9pt}{}}^{\!\!\!b}\right)}$&
\cr
height3pt
&\omit&
&\omit&
&\omit&
\cr
\noalign{\hrule}
height3pt
&\omit&
&\omit&
&\omit&
\cr
&\hfil $\c\Z_7\c$&
&$\eqalign{(1,2,-3)/7}$&
&$\eqalign{
{\rm diag}\left(\alpha^{2}{\bf I_{4}\raisebox{9pt}{}}^{\!\!\!a}, 
\alpha^{- 2}{\bf I_{\bar 4}\raisebox{9pt}{}}^{\!\!\!a}, 
\alpha^{4}{\bf I_{4}\raisebox{9pt}{}}^{\!\!\!b}, 
\alpha^{- 4}{\bf I_{\bar 4}\raisebox{9pt}{}}^{\!\!\!b}, 
\alpha^{6}{\bf I_{4}\raisebox{9pt}{}}^{\!\!\!c}, 
\alpha^{- 6}{\bf I_{\bar 4}\raisebox{9pt}{}}^{\!\!\!c},
{\bf I_{8}\raisebox{9pt}{}}^{\!\!\!d}\right)}$&
\cr 
height3pt 
&\omit&
&\omit&
&\omit&
\cr
\noalign{\hrule}
height3pt 
&\omit& 
&\omit&
&\omit&
\cr
&\hfil $\c\Z_6$\c&
&$\eqalign{(1,1,-2)/6}$&
&$\eqalign{
{\rm diag}\left(\alpha \,{\bf I_{6}\raisebox{9pt}{}}^{\!\!\!a}, 
\alpha^{- 1}{\bf I_{\bar 6}\raisebox{9pt}{}}^{\!\!\!a}, 
\alpha^{5}{\bf I_{6}\raisebox{9pt}{}}^{\!\!\!b}, 
\alpha^{- 5}{\bf I_{\bar 6}\raisebox{9pt}{}}^{\!\!\!b}, 
\alpha^{3}{\bf I_{4}\raisebox{9pt}{}}^{\!\!\!c}, 
\alpha^{- 3}{\bf I_{\bar 4}\raisebox{9pt}{}}^{\!\!\!c}\right)}$&
\cr
height3pt 
&\omit& 
&\omit&
&\omit&
\cr
\noalign{\hrule} 
height3pt 
&\omit& 
&\omit&
&\omit&
\cr
&\hfil $\c\Z_6^\prime\c$&
&$\eqalign{(1,-3,2)/6}$&
&$\eqalign{
{\rm diag}\left(\alpha \,{\bf I_{4}\raisebox{9pt}{}}^{\!\!\!a}, 
\alpha^{- 1}{\bf I_{\bar 4}\raisebox{9pt}{}}^{\!\!\!a}, 
\alpha^{5}{\bf I_{4}\raisebox{9pt}{}}^{\!\!\!b}, 
\alpha^{- 5}{\bf I_{\bar 4}\raisebox{9pt}{}}^{\!\!\!b}, 
\alpha^{3}{\bf I_{8}\raisebox{9pt}{}}^{\!\!\!c}, 
\alpha^{- 3}{\bf I_{\bar 8}\raisebox{9pt}{}}^{\!\!\!c}\right)}$&
\cr 
height3pt 
&\omit& 
&\omit&
&\omit&
\cr
\noalign{\hrule} 
height3pt 
&\omit& 
&\omit&
&\omit&
\cr
&\hfil $\c\Z_{12}\c$&
&$\eqalign{(1,-5,4)/12 \cr \cr}$&
&$\eqalign{
{\rm diag}\left(\alpha^{- 1}{\bf I_{3}\raisebox{9pt}{}}^{\!\!\!a}, 
\alpha \,{\bf I_{\bar 3}\raisebox{9pt}{}}^{\!\!\!a}, 
\alpha^{5}{\bf I_{3}\raisebox{9pt}{}}^{\!\!\!b}, 
\alpha^{- 5}{\bf I_{\bar 3}\raisebox{9pt}{}}^{\!\!\!b}, 
\alpha^{- 7}{\bf I_{3}\raisebox{9pt}{}}^{\!\!\!c}, 
\alpha^{7}{\bf I_{\bar 3}\raisebox{9pt}{}}^{\!\!\!c}, 
\alpha^{11}{\bf I_{3}\raisebox{9pt}{}}^{\!\!\!d}, 
\alpha^{- 11}{\bf I_{\bar 3}\raisebox{9pt}{}}^{\!\!\!d},
\right.\cr\left.\hspace{29pt}
\alpha^{3}{\bf I_{2}\raisebox{9pt}{}}^{\!\!\!e}, 
\alpha^{- 3}{\bf I_{\bar 2}\raisebox{9pt}{}}^{\!\!\!e},
\alpha^{9}{\bf I_{2}\raisebox{9pt}{}}^{\!\!\!f}, 
\alpha^{- 9}{\bf I_{\bar 2}\raisebox{9pt}{}}^{\!\!\!f}\right)}$&
\cr 
height3pt 
&\omit& 
&\omit&
&\omit&
\cr
}
\hrule height 1.1pt}
$$
}
\noindent
\caption{Twist vectors and matrices for the $\Z_N$ models.
Latin letters refer to the various factors of the gauge group 
reported in table 3, ${\bf I_{\rho}}$ indicates the identity in 
the representation $\rho$, and $\alpha = e^{\frac {\pi}N i}$.}
\label{vN}
\end{table}

\vskip 10pt 
\noindent 
{\bf $\Z_N \times \Z_M$ orientifolds}
\vskip 5pt 

\noindent
The $\Z_N \times \Z_M$ action is generated by the elements
$\theta = exp (2 \pi i v_i J_i)$,  $\omega = exp (2 \pi i w_i J_i)$,
associated to the twist vectors $v = (v_1,v_2,v_3)$ and 
$w = (w_1,w_2,w_3)$. The matrices $\gamma_k$ and $\delta_l$ representing 
the twists $\theta^k$ and $\omega^l$ on open strings can be chosen such 
that $\gamma_{k,9} = (\gamma)^k$, $\gamma_{k,5} = (-\gamma)^k$,
$\delta_{l,9} = \delta_{l,5} = (\delta)^l$ \cite{z}. 
The $\Z_N \times \Z_M$ actions in table \ref{vNM} lead to consistent 
model with cancelled tadpoles. 

\begin{table}[h]
\vbox{
$$\vbox{\offinterlineskip
\hrule height 1.1pt
\halign{&\vrule width 1.1pt#
&\strut\quad#\hfil\quad&
\vrule#
&\strut\quad#\hfil\quad&
\vrule#
&\strut\quad#\hfil\quad&
\vrule width 1.1pt#\cr
height3pt
&\omit&
&\omit&
&\omit&
\cr
&\hfil $G$ &
&\hfil $v,w$&
&\hfil $\gamma,\delta$ &
\cr
height3pt
&\omit&
&\omit&
&\omit&
\cr
\noalign{\hrule height 1.1pt}
height3pt
&\omit& 
&\omit&
&\omit&
\cr
&\hfil $\c\Z_3 \times \Z_3\c$&
&$\eqalign{(1,-1,0)/3,\smallskip\ \cr (0,1,-1)/3}$&
&$\eqalign{
{\rm diag}\left(\alpha^{2}{\bf I_{4}\raisebox{9pt}{}}^{\!\!\!a}, 
\alpha^{- 2}{\bf I_{\bar 4}\raisebox{9pt}{}}^{\!\!\!a},
\alpha^{4}{\bf I_{4}\raisebox{9pt}{}}^{\!\!\!b}, 
\alpha^{- 4}{\bf I_{\bar 4}\raisebox{9pt}{}}^{\!\!\!b},
{\bf I_{4}\raisebox{9pt}{}}^{\!\!\!c}, 
{\bf I_{\bar 4}\raisebox{9pt}{}}^{\!\!\!c},
{\bf I_{8}\raisebox{9pt}{}}^{\!\!\!d}\right),\smallskip\ \cr
{\rm diag}\left({\bf I_{4}\raisebox{9pt}{}}^{\!\!\!a}, 
{\bf I_{\bar 4}\raisebox{9pt}{}}^{\!\!\!a},
\beta^{2}{\bf I_{4}\raisebox{9pt}{}}^{\!\!\!b}, 
\beta^{- 2}{\bf I_{\bar 4}\raisebox{9pt}{}}^{\!\!\!b},
\beta^{4}{\bf I_{4}\raisebox{9pt}{}}^{\!\!\!c}, 
\beta^{- 4}{\bf I_{\bar 4}\raisebox{9pt}{}}^{\!\!\!c},
{\bf I_{8}\raisebox{9pt}{}}^{\!\!\!d}\right)}$&
\cr 
height3pt 
&\omit& 
&\omit&
&\omit&
\cr
\noalign{\hrule} 
height3pt 
&\omit& 
&\omit&
&\omit&
\cr
&\hfil $\c\Z_3 \times \Z_6\c$&
&$\eqalign{(1,0,-1)/3, \cr \medskip\ \cr (1,-1,0)/6 \cr \cr}$&
&$\eqalign{
{\rm diag}\left(\alpha^{4}{\bf I_{2}\raisebox{9pt}{}}^{\!\!\!a}, 
\alpha^{- 4}{\bf I_{\bar 2}\raisebox{9pt}{}}^{\!\!\!a},
{\bf I_{2}\raisebox{9pt}{}}^{\!\!\!b}, 
{\bf I_{\bar 2}\raisebox{9pt}{}}^{\!\!\!b},
\alpha^{2}{\bf I_{2}\raisebox{9pt}{}}^{\!\!\!c}, 
\alpha^{- 2}{\bf I_{\bar 2}\raisebox{9pt}{}}^{\!\!\!c},
{\bf I_{2}\raisebox{9pt}{}}^{\!\!\!d}, 
{\bf I_{\bar 2}\raisebox{9pt}{}}^{\!\!\!d},
\right.\cr\left.\hspace{29pt}
\alpha^{2}{\bf I_{2}\raisebox{9pt}{}}^{\!\!\!e}, 
\alpha^{- 2}{\bf I_{\bar 2}\raisebox{9pt}{}}^{\!\!\!e},
\alpha^{4}{\bf I_{2}\raisebox{9pt}{}}^{\!\!\!f}, 
\alpha^{- 4}{\bf I_{\bar 2}\raisebox{9pt}{}}^{\!\!\!f},
{\bf I_{4}\raisebox{9pt}{}}^{\!\!\!g},
{\bf I_{\bar 4}\raisebox{9pt}{}}^{\!\!\!g}\right), \smallskip\ \cr
{\rm diag}\left(\beta^{- 1}{\bf I_{2}\raisebox{9pt}{}}^{\!\!\!a}, 
\beta \,{\bf I_{\bar 2}\raisebox{9pt}{}}^{\!\!\!a},
\beta^{- 1}{\bf I_{2}\raisebox{9pt}{}}^{\!\!\!b}, 
\beta \,{\bf I_{\bar 2}\raisebox{9pt}{}}^{\!\!\!b},
\beta^{- 5}{\bf I_{2}\raisebox{9pt}{}}^{\!\!\!c}, 
\beta^{5}{\bf I_{\bar 2}\raisebox{9pt}{}}^{\!\!\!c},
\beta^{- 5}{\bf I_{2}\raisebox{9pt}{}}^{\!\!\!d}, 
\beta^{5}{\bf I_{\bar 2}\raisebox{9pt}{}}^{\!\!\!d},
\right.\cr\left.\hspace{29pt}
\beta^{3}{\bf I_{2}\raisebox{9pt}{}}^{\!\!\!e}, 
\beta^{- 3}{\bf I_{\bar 2}\raisebox{9pt}{}}^{\!\!\!e},
\beta^{3}{\bf I_{2}\raisebox{9pt}{}}^{\!\!\!f}, 
\beta^{- 3}{\bf I_{\bar 2}\raisebox{9pt}{}}^{\!\!\!f},
\beta^{3}{\bf I_{4}\raisebox{9pt}{}}^{\!\!\!g},
\beta^{- 3}{\bf I_{\bar 4}\raisebox{9pt}{}}^{\!\!\!g}\right)}$&
\cr 
height3pt 
&\omit&
&\omit&
&\omit&
\cr
}
\hrule height 1.1pt}
$$
}
\caption{Twist vectors and matrices for the $\Z_N\times \Z_M$ models.
Latin letters refer to the various factors of the gauge group 
reported in table 4, ${\bf I_{\rho}}$ indicates the identity in the 
representation $\rho$, and $\alpha = e^{\frac {\pi}N i}$, 
$\beta = e^{\frac {\pi}M i}$.}
\label{vNM}
\end{table}

\vskip 5pt

The spectra of these orientifold models have been analysed in 
\cite{abpss,ks1,ks2,afiv,z}, but there are some minor discrepancies
about the charged states content, aside those arising from
some arbitrariness in the choice of the twist matrices in tables
\ref{vN} and \ref{vNM}. We fix these spectra by comparing the standard 
field-theory computation and our string theory computation of the anomaly. 
The result are reported in the tables of next section.


\section{Gauge and gravitational anomalies}

The models we are considering have in general non-vanishing 
gauge/gravitational one-loop anomalies, which are expected 
to be cancelled by an inflow of anomaly through the GS mechanism. 
The latter can be deduced only from a direct 
string theory computation. It has been shown in 
\cite{ss2} that the total anomaly vanishes for any orientifold model with
cancelled tadpoles, as a consequence of the cancellation of two equal and 
opposite contributions which can be identified with the total one-loop 
anomaly and the total tree-level inflow respectively. The central result 
of \cite{ss2} is that the anomaly and inflow polynomials are both given 
by the same one-loop partition function in the RR odd spin-structure, in 
external gauge and gravitational backgrounds~\footnote{In order to compute the 
anomaly polynomial and not the anomaly itself, one has to go in two 
dimensions higher and omit the bosonic zero modes.}. Using this strategy, 
one can analyze in detail the pattern of anomaly cancellation and deduce
by factorization (up to overall signs and trivial rescaling of the fields) the 
relevant CP-odd couplings in the low-energy action. 

Recall that the contribution to the one-loop anomaly polynomial from a chiral 
spinor in the representation $\rho$ of the gauge group, in a gauge and 
gravitational background with curvatures $F$ and $R$, involves the 
Chern-character of the gauge bundle and the Roof-genus of the tangent bundle, 
and is given by the famous formula~\footnote{The corresponding anomaly is 
given by the Wess-Zumino descent $A = 2 \pi i \int I^{(1)}$.}
$$
I_{1/2}^\rho = {\rm ch}_{\bf \rho} (F) \, \widehat A (R) \;.
$$

It is convenient to decompose all the representations of the $U(n)$ and 
$SO(n)$ factors as tensor products of two fundamental representations 
associated to the end-points of open strings. Correspondingly, the Chern 
character appearing in the anomaly decompose as products of Chern characters 
$c(F)$ in the fundamental representation,
$$
c(F) = {\rm ch}_{\bf n} (F) = {\rm tr}_{\bf n} [e^{iF/2\pi}] \;.
$$
The adjoint representations of $U(n)$ and $SO(n)$ do not contribute,
and for the antisymmetric representation of $U(n)$ one gets the following
decomposition:
\be
{\rm ch}_{\bf \frac {n(n - 1)}2}(F) = \frac 12 
\left[c(F)^2 - c(2F)\right] \;.
\label{decomp}
\ee
From a string theory point of view, the polynomial associated to both
the anomaly and the inflow is given by the total one-loop partition 
function in the odd spin-structure. Only the annulus and the M\"obius strip 
amplitudes contribute. The Klein bottle contribution vanish because of 
the impossibility of having the correct number of fermionic zero modes
inserted. This reflects the fact that there are no pure gravitational
anomalies in four dimensions and only charged states can contribute to the 
anomaly. Summing over all the D-brane sectors, one gets therefore 
\be
I = \raisebox{2pt}{$\displaystyle{\sum_{\alpha,\beta=9,5}}$} 
I_{\cal A}^{\alpha \beta} 
+ \raisebox{2pt}{$\displaystyle{\sum_{\alpha=9,5}}$}
I_{\cal M}^{\alpha} \;. 
\label{totinf}
\ee
We shall see in the following that the annulus partition function
produces naturally a contribution proportional to $c(F)^2$, the 
M\"obius strip giving instead $c(2F)$. Comparing with (\ref{decomp}), this 
observation allows to reconstruct the anomalous charged 
matter spectrum of each model from the string theory expression for 
the anomaly polynomial.

\subsection{$\Z_N$ orientifolds}

In the operatorial formalism, the relevant odd spin-structure amplitudes 
to compute on the annulus and the M\"obius strip are encoded in the 
following partition functions:
\bea
\a\a I_{\cal A} = \frac{1}{4N}\sum_{k=0}^{N-1}\,
{\rm Tr}_{R}\,[\theta^k\,(-1)^F\,e^{-tH(F,R)}] \label{AN} \;,\\ 
\a\a I_{\cal M} =\frac{1}{4N}\sum_{k=0}^{N-1}\, 
{\rm Tr}_R\,[\Omega\,\theta^k\,(-1)^F\,e^{-tH(F,R)}] \label{MN}  \;,
\eea
where a sum over Chan-Paton indices is understood.
Their evaluation is rather straightforward. 
Being topological indices,
only massless modes contribute, massive states cancelling by supersymmetry.
It is convenient to define, in each compact plane $i=1,2,3$ and each
$k$-twisted sector, $k=1,...,N-1$, the quantities 
$$
s^i_{k} = 2 \sin \pi k v_i \;,\;\;
c^i_{k} = 2 \cos \pi k v_i \;. 
$$
The number of $k$-fixed-points, that is those points which are fixed under 
$k$-twists, is $N_k^i = (s_k^i)^2$ ``per plane'', and in total one  
has $N_k = N_k^1 N_k^2 N_k^3$ fixed-points in the whole compact space 
and $N_k^3$ in the third plane, where the D5-branes wrap. Also, 
it is natural to introduce the modified $\Z_N$ Chern-character 
\be
{\rm ch}_{k} (F) = {\rm tr} [\gamma_k \,e^{i F/2\pi}] \;,
\ee
where the trace is in the Chan-Paton representation. 

The untwisted $k=0$ sector vanish for all the surfaces and D-brane sectors, 
because of the presence of unsoaked fermionic zero modes in the internal 
directions, and we restrict therefore to $k\neq 0$. 
On the annulus, a pair of bosons and fermions with Neumann Neumann (NN)
boundary conditions in the $i$-th compact plane contribute 
$(s_k^i)^{-2}$ and $s_k^i$ respectively, whereas a pair of bosons and 
fermions with Dirichlet Dirichlet (DD) boundary conditions give $1$ and 
$s_{k}^i$. No fields with ND boundary conditions contribute,
having a half-integer mode expansion and correspondingly no zero-energy
states. The fields in the non-compact directions give instead the contribution 
$i \,{\rm ch}_k (F_\alpha)\,{\rm ch}_k(F_\beta)\,\widehat{A}(R)$ in 
the $\alpha \beta$ sector. Taking into account 
the number of fixed-points, $N_k$ in the 99 sector 
and $N_k^3$ in the 55 and 95 sectors, one finds
\be
I_{\cal A}^{\alpha \beta} = \frac{i}{4N}\,\sum_{k=1}^{N-1}
C_k^{\alpha \beta}\,{\rm ch}_k (F_\alpha)\,
{\rm ch}_k (F_\beta)\,\widehat A(R) \;, \label{ZAN}
\ee
where
\be
C_k^{\alpha \beta}= \left\{\!\!
\begin{array}{l}
s_k^1 s_k^2 s_k^3 \;,\;\; \alpha = \beta \medskip\ \\
s_k^3 \;,\;\; \alpha \neq \beta \;.
\end{array} \right.
\ee
On the M\"obius strip a pair of N bosons and fermions in the $i$-th 
compact plane give the contributions $(s_k^i)^{-2}$ and 
$s_k^i$ respectively, whereas a pair of D bosons and 
fermions give respectively $1$ and $c_k^i$. The fields in 
the non-compact directions give instead the contribution 
$i\,{\rm ch}_{2k}(2F_\alpha)\,\widehat{A}(R)$ 
in the $\alpha$ sector. Taking again into account the number of 
fixed-points, $N_k$ in the 9 sector and $N_k^3$ in the 5 sector,
one finds 
\be
I_{\cal M}^{\alpha} = - \frac{i}{4N}\,\sum_{k=1}^{N-1}
C_k^{\alpha}\,{\rm ch}_{2k}(2F_\alpha)\,
\widehat A(R) \;, \label{ZMN}
\ee
where
\be
C_k^{\alpha}= \left\{\!\!
\begin{array}{l}
s_k^1 s_k^2 s_k^3 \;,\;\; \alpha = 9  \medskip\ \\
c_k^1 c_k^2 s_k^3 \;,\;\; \alpha = 5\;.
\end{array} \right.
\ee
One can check case by case that the two trigonometric factors 
in $I_{\cal M}^{9,5}$ give always identical contributions, making 
manifest the $9 \leftrightarrow 5$ symmetry implied by T-duality.

The massless open string spectrum which is in agreement with the one-loop
anomaly (\ref{totinf}) is reported in table \ref{csN}.

\begin{table}[h]
\vbox{
$$\vbox{\offinterlineskip
\hrule height 1.1pt
\halign{&\vrule width 1.1pt#
&\strut\quad#\hfil\quad&
\vrule#
&\strut\quad#\hfil\quad&
\vrule#
&\strut\quad#\hfil\quad&
\vrule#
&\strut\quad#\hfil\quad&
\vrule width 1.1pt#\cr
height3pt
&\omit&
&\omit&
&\omit&
&\omit&
\cr
&\hfil $G$ &
&\hfil Gauge Group &
&\hfil 99/55 Matter &
&\hfil 95 Matter &
\cr
height3pt
&\omit&
&\omit&
&\omit&
&\omit&
\cr
\noalign{\hrule height 1.1pt}
height3pt
&\omit&
&\omit&
&\omit&
&\omit&
\cr
&\hfil $\c\Z_3\c$&
&$\eqalign{U(12) \times SO(8)}$&
&$\eqalign{
3 {\bf (12,8)},\; 3 {\bf (\bar{66},1)}}$&
&\hfil -&
\cr
height3pt
&\omit&
&\omit&
&\omit&
&\omit&
\cr
\noalign{\hrule}
height3pt
&\omit&
&\omit&
&\omit&
&\omit&
\cr
&\hfil $\c\Z_7\c$&
&$\eqalign{U(4)^3 \times SO(8) \cr \cr}$&
&$\eqalign{ 
{\bf (\underline{4,1,1},8)},\;{\bf (\underline{6,1,1},1)}\cr
{\bf (\underline{\bar 4,\bar 4,1},1)},\;{\bf (\underline{\bar 4,1,4},1)}}$&
&\hfil -&
\cr 
height3pt 
&\omit&
&\omit& 
&\omit& 
&\omit&
\cr
\noalign{\hrule}
height3pt 
&\omit&
&\omit& 
&\omit& 
&\omit&
\cr
&\hfil $\c\Z_6\c$&
&$\eqalign{\left(U(6)^2 \times U(4)\right)^2 \cr \cr \cr \cr \cr \cr}$&
&$\eqalign{
2 {\bf (15,1,1)},\; 2 {\bf (1,\bar{15},1)}\cr
2 {\bf (\bar 6,1,4)},\; 2 {\bf (1,6,\bar 4)}\cr
{\bf (\bar 6,1,\bar 4)},\; {\bf (1,6,4)}\cr
{\bf (6,\bar 6,1)}\cr
\cr
\cr}$&
&$\eqalign{
{\bf (6,1,1;6,1,1)}\cr 
{\bf (1,\bar 6,1;1,\bar 6,1)}\cr
{\bf (1,6,1;1,1,\bar 4)}\cr
{\bf (1,1,\bar 4;1,6,1)}\cr
{\bf (\bar 6,1,1;1,1,4)}\cr
{\bf (1,1,4;\bar 6,1,1)}}$&
\cr
height3pt 
&\omit&
&\omit& 
&\omit& 
&\omit&
\cr
\noalign{\hrule} 
height3pt 
&\omit&
&\omit& 
&\omit& 
&\omit&
\cr
&\hfil $\c\Z_6^\prime\c$&
&$\eqalign{\left(U(4)^2 \times U(8)\right)^2 \cr \cr \cr \cr \cr \cr}$&
&$\eqalign{
{\bf (4,1,8)},\; {\bf (\bar 4,1,8)}\cr
{\bf (1,4,\bar 8)},\; {\bf (1,\bar 4,\bar 8)}\cr
{\bf (1,1,28)},\; {\bf (1,1,\bar{28})}\cr
{\bf (6,1,1)},\; {\bf (1,\bar 6,1)}\cr
{\bf (4,4,1)},\; {\bf (\bar 4,\bar 4,1)}\cr
{\bf (\bar 4,4,1)}}$&
&$\eqalign{
{\bf (\bar 4,1,1;\bar 4,1,1)}\cr 
{\bf (1,4,1;1,4,1)}\cr
{\bf (1,\bar 4,1;1,1,8)}\cr
{\bf (1,1,8;1,\bar 4,1)}\cr
{\bf (4,1,1;1,1,\bar 8)}\cr
{\bf (1,1,\bar 8;4,1,1)}}$&
\cr 
height3pt 
&\omit&
&\omit& 
&\omit& 
&\omit&
\cr
\noalign{\hrule} 
height3pt 
&\omit&
&\omit& 
&\omit& 
&\omit&
\cr
&\hfil $\c\Z_{12}\c$&
&$\eqalign{\left(U(3)^4 \times U(2)^2\right)^2 
\cr \cr \cr \cr \cr \cr \cr \cr}$&
&$\eqalign{
{\bf (\underline{\bar 3,1},\underline{\bar 3,1},1,1)}\cr
{\bf (\underline{\bar 3_A,1,1,1},1,1)}\cr
{\bf (\underline{3,1},1,1,\underline{2,1})}\cr
{\bf (1,1,\underline{3,1},\underline{\bar 2,1})}\cr
{\bf (1,1,1,3,2,1)}\cr
{\bf (1,3,1,1,\bar 2,1)}\cr
{\bf (1,1,3,1,1,2)}\cr
{\bf (3,1,1,1,1,\bar 2)}}$&
&$\eqalign{
{\bf (\bar 3,1^5;1,\bar 3,1^4)}\cr 
{\bf (1^2,\bar 3,1^3;1^3,\bar 3,1^2)}\cr 
{\bf (3,1^5;1^4,\bar 2,1)}\cr 
{\bf (1^2,3,1^3;1^4,2,1)}\cr 
{\bf (1,3,1^4;1^5,\bar 2)}\cr 
{\bf (1^3,3,1^2;1^5,2)}\cr
+ \, (9 \leftrightarrow 5)
\cr \cr}$&
\cr
height3pt 
&\omit&
&\omit& 
&\omit& 
&\omit&
\cr
}
\hrule height 1.1pt}
$$
}
\caption{Massless open string spectrum for $Z_N$ models. The underlined
bar means that all the cyclic permutations have to be considered.
Our conventions for $U(1)$ charges are such that the ${\bf n}$ and 
${\bf \bar n}$ of $U(n)$ carry $\pm 1$ charge with respect to the 
corresponding $U(1)$.}
\label{csN}
\end{table}

\noindent
As a consequence of tadpole cancellation, all irreducible non-Abelian 
gauge anomalies vanish. Therefore, only mixed $U(1)$-gauge and 
$U(1)$-gravitational anomalies arise. The former gets contribution
only from the annulus, whereas for the latter both the annulus and the 
M\"obius strip contribute. The explicit form of the total anomaly is 
\bea
\a\a I = \frac{1}{4N(2\pi)^3}\,\sum_{k=1}^{N-1} 
\left\{\raisebox{2pt}{$\displaystyle{\sum_{\alpha,\beta=9,5}}$}
C_k^{\alpha \beta}\,{\rm tr}(\gamma_k F_\alpha)\,
{\rm tr}(\gamma_k F_\beta^2) \right. \label{explN} \\ \a\a \hspace{100pt}
+ \frac 13 \left[\raisebox{2pt}{$\displaystyle{\sum_{\alpha,\beta=9,5}}$}
C_k^{\alpha \beta}\,{\rm tr}(\gamma_{k,\alpha})\,{\rm tr}(\gamma_k F_\beta^3) 
- 4 \raisebox{2pt}{$\displaystyle{\sum_{\alpha=9,5}}$}
C_k^{\alpha}\,{\rm tr}(\gamma_{2k} F_\alpha^3)\right] \nn \\
\a\a \hspace{100pt} \left. - \frac{1}{24}\left[ 
\raisebox{2pt}{$\displaystyle{\sum_{\alpha,\beta=9,5}}$}
C_k^{\alpha \beta}\,{\rm tr} (\gamma_{k,\alpha})\,{\rm tr}(\gamma_k F_\beta) 
- \raisebox{2pt}{$\displaystyle{\sum_{\alpha=9,5}}$}
C_k^{\alpha}\,{\rm tr}(\gamma_{2k} F_\alpha) \right] 
{\rm tr} R^2 \right\} \nn \;.
\eea
The cancellation of all the non-Abelian gauge anomalies (second raw of
(\ref{explN})) requires the non-trivial identity
\be
\sum_{k=1}^{N-1} C_k^{\alpha}\,
{\rm tr}(\gamma_{2k} F_\alpha^{2n+1}) = \frac 14
\sum_{k=1}^{N-1} \raisebox{2pt}{$\displaystyle{\sum_{\beta=9,5}}$}
C_k^{\alpha \beta}\,{\rm tr}(\gamma_{k,\beta})\,
{\rm tr}(\gamma_k F_\alpha^{2n+1}) \;. \label{condN}
\ee
This is indeed implied by the tadpole-cancellation conditions \cite{abiu}.
Using (\ref{condN}), the total mixed $U(1)$-gauge/gravitational anomaly
is finally found to be given by the simple expression
\be
I = \frac{1}{4N(2\pi)^3}\,\sum_{k=1}^{N-1} 
\raisebox{2pt}{$\displaystyle{\sum_{\alpha,\beta=9,5}}$} 
C_k^{\alpha \beta}\,{\rm tr}(\gamma_k F_\alpha)\,
\left({\rm tr}(\gamma_k F_\beta^2) 
- \frac 1{32}\,{\rm tr}(\gamma_{k,\beta})\, 
{\rm tr} R^2 \right) \label{simplN}\;.
\ee
This result is in agreement with equations (3.5) and (3.8) of \cite{iru1},
corresponding to the first and second term respectively.

\subsection{$\Z_N \times \Z_M$ orientifolds}

The relevant odd spin-structure partition functions to be computed are 
now given by the following expressions
\bea
\a\a I_{\cal A} = \frac{1}{4NM} \sum_{k=0}^{N-1}\sum_{l=0}^{M-1}\,
{\rm Tr}_{R}\,[\theta^k\, \omega^l \, (-1)^F\,e^{-tH(F,R)}] \label{ANM} \;,\\ 
\a\a I_{\cal M} =\frac{1}{4NM}\sum_{k=0}^{N-1}\sum_{l=0}^{M-1}\, 
{\rm Tr}_R\,[\Omega\,\theta^k\,\omega^l \,(-1)^F\,e^{-tH(F,R)}] 
\label{MNM}  \;.
\eea
Their evaluation is again quite straightforward and the total inflow 
is given by (\ref{totinf}). As before, it is very convenient to define, in 
each compact plane $i=1,2,3$ and each $(k,l)$-twisted sector, 
$k=0,...,N-1$, $l=0,...,M-1$, $(k,l) \neq (0,0)$, the quantities
$$
s^i_{k,l} = 2 \sin \pi (k v_i + l w_i) \;,\;\;
c^i_{k,l} = 2 \cos \pi (k v_i + l w_i) \;. 
$$
The number of $(k,l)$-fixed-points is $N_{k,l}^i = (s_{k,l}^i)^2$ 
``per plane'', and in total there are 
$N_{k,l} = N_{k,l}^1 N_{k,l}^2 N_{k,l}^3$ 
fixed-points in the whole compact space and $N_{k,l}^3$ 
in the third plane, where the D5-branes wrap. Also, the appropriate
$\Z_N \times \Z_M$ Chern-character is
\be
{\rm ch}_{k,l} (F) = {\rm tr} [\gamma_k\,\delta_l\,e^{i F/2\pi}] \;.
\ee

The evaluation of (\ref{ANM}) and (\ref{MNM}) proceeds along
the lines of the previous case. For the annulus, one gets
\be
I_{\cal A}^{\alpha \beta} = \frac{i}{4NM}\,
\sum_{k=0}^{N-1}\sum_{l=0}^{M-1}
C_{k,l}^{\alpha \beta}\,{\rm ch}_{k,l}(F_\alpha)\,
{\rm ch}_{k,l}(F_\beta)\,\widehat A(R) \;, \label{ZANM}
\ee
where
\be
C_{k,l}^{\alpha \beta} = \left\{\!\!
\begin{array}{l}
s_{k,l}^1 s_{k,l}^2 s_{k,l}^3 \;,\;\; \alpha = \beta \medskip\ \\
s_{k,l}^3 \;,\;\; \alpha \neq \beta
\end{array} \right. \; \bigskip\
\ee
and for the M\"obius strip 
\be
I_{\cal M}^{\alpha} = - \frac{i}{4NM}\,\sum_{k=0}^{N-1}\sum_{l=0}^{M-1}
C_{k,l}^{\alpha}\,{\rm ch}_{2k,2l}(2F_\alpha)\,\widehat A(R) 
\;, \label{ZMNM}
\ee
where
\be
C_{k,l}^{\alpha} = \left\{\!\!
\begin{array}{l}
s_{k,l}^1 s_{k,l}^2 s_{k,l}^3 \;,\;\; \alpha = 9 \medskip\ \\
c_{k,l}^1 c_{k,l}^2 s_{k,l}^3 \;,\;\; \alpha = 5 \;.
\end{array} \right. \bigskip\
\ee
Again, one can check that the two trigonometric factors in 
$Z_{\cal M}^{9,5}$ give identical contributions.

The massless open string spectrum corresponding to the one-loop
anomaly (\ref{totinf}) is reported in table \ref{csNM}.

\begin{table}[h]
\vbox{
$$\vbox{\offinterlineskip
\hrule height 1.1pt
\halign{&\vrule width 1.1pt#
&\strut\quad#\hfil\quad&
\vrule#
&\strut\quad#\hfil\quad&
\vrule#
&\strut\quad#\hfil\quad&
\vrule#
&\strut\quad#\hfil\quad&
\vrule width 1.1pt#\cr
height3pt
&\omit&
&\omit&
&\omit&
&\omit&
\cr
&\hfil $G$ &
&\hfil Gauge Group &
&\hfil 99/55 Matter &
&\hfil 95 Matter &
\cr
height3pt
&\omit&
&\omit&
&\omit&
&\omit&
\cr
\noalign{\hrule height 1.1pt}
height3pt
&\omit&
&\omit&
&\omit&
&\omit&
\cr
&\hfil $\c\Z_3 \times \Z_3\c$&
&$\eqalign{U(4)^3 \times SO(8)}$&
&$\eqalign{
{\bf (\underline{4,1,1},8)},\; {\bf (\underline{\bar 6,1,1},1)},\cr
{\bf (\underline{\bar 4,\bar 4,1},1)}}$&
&\hfil -&
\cr
height3pt
&\omit&
&\omit&
&\omit&
&\omit&
\cr
\noalign{\hrule}
height3pt 
&\omit&
&\omit& 
&\omit& 
&\omit&
\cr
&\hfil $\c\Z_3 \times \Z_6\c$&
&$\eqalign{\left(U(2)^6 \times U(4)\right)^2 
\cr \cr \cr \cr \cr \cr \cr \cr \cr}$&
&$\eqalign{
{\bf (1,1_A,1^5)},\; {\bf (1^3,\bar 1_A,1^3)}\cr
{\bf (2,\bar 2,1^5)},\; {\bf (\bar 2,\bar 2,1^5)}\cr
{\bf (1^2,\bar 2,2,1^3)},\;{\bf (1^2,2,2,1^3)}\cr
{\bf (1^4,2,\bar 2,1)}\cr
{\bf (\bar 2,1^3,\bar 2,1^2)},\;{\bf (1,2,1^2,2,1^2)}\cr
{\bf (1^2,\bar 2,1,\bar 2,1^2)},\;{\bf (2,1^4,2,1)}\cr
{\bf (1^3,\bar 2,1,\bar 2,1)},\;{\bf (1^2,2,1^2,2,1)}\cr
{\bf (1,\bar 2,1^4,\bar 4)},\;{\bf (1^3,2,1^2,4)}\cr
{\bf (1^4,\bar 2,1,4)},\;{\bf (1^5,2,\bar 4)}\cr}$&
&$\eqalign{
{\bf (2,1^6;1,\bar 2,1^5)}\cr 
{\bf (1^2,\bar 2,1^4;1^3,2,1^3)}\cr 
{\bf (1^4,2,1^2;1^5,\bar 2,1)}\cr 
{\bf (1^5,2,1;1^6,\bar 4)}\cr 
{\bf (1^4,\bar 2,1^2;1^6, 4)}\cr 
+ \, (9 \leftrightarrow 5)\cr
\cr \cr \cr}$&
\cr
height3pt 
&\omit&
&\omit& 
&\omit& 
&\omit&
\cr
}
\hrule height 1.1pt}
$$
}
\caption{Massless open string spectrum for $\Z_N \times \Z_M$ models.}
\label{csNM}
\end{table}

\noindent
As before, all irreducible non-Abelian gauge anomalies vanish due to a 
cancellation between the annulus and the M\"obius strip contributions, 
and only mixed $U(1)$-gauge and $U(1)$-gravitational anomalies arise. 
More precisely, the condition imposed by the vanishing of all irreducible 
non-Abelian gauge anomalies is
\be
\sum_{k=0}^{N-1} \sum_{l=0}^{M-1} C_{k,l}^{\alpha}\,
{\rm tr}(\gamma_{2k} \delta_{2l} F_\alpha^{2n+1}) = \frac 14
\sum_{k=0}^{N-1} \sum_{l=0}^{M-1}
\raisebox{2pt}{$\displaystyle{\sum_{\beta=9,5}}$}
C_{k,l}^{\alpha \beta}\,{\rm tr}(\gamma_{k,\beta} \delta_{l,\beta})\,
{\rm tr}(\gamma_k \delta_l F_\alpha^{2n+1}) \;. \label{condNM}
\ee
Using this condition, the total $U(1)$-gauge/gravitational anomaly takes 
finally the following simple form:
\be
\hspace{-0pt}
I = \frac{1}{4NM(2\pi)^3} \sum_{k=0}^{N-1}\sum_{l=0}^{M-1} 
\raisebox{2pt}{$\displaystyle{\sum_{\alpha,\beta=9,5}}$} 
C_{k,l}^{\alpha \beta}\,{\rm tr}(\gamma_k \delta_l F_\alpha)
\left({\rm tr}(\gamma_k \delta_l F_\beta^2)
- \frac 1{32} {\rm tr}(\gamma_{k,\beta} \delta_{l,\beta})\, 
{\rm tr} R^2 \right) . \hspace{-8pt} \label{simplNM}
\ee


\section{Factorization}

Having computed all the inflows, the anomalous couplings to RR fields 
can be obtained by factorization, in the spirit of \cite{ss1} (see
also \cite{ss3}). 
In order to do so, one needs a precise knowledge of the massless 
closed string spectra, which have been studied in detail in 
\cite{afiv,klein}. Analyzing (\ref{ZAN})-(\ref{ZMN}) 
and (\ref{ZANM})-(\ref{ZMNM}) in the transverse channel, it is then 
possible to identify, in most cases, which of these states mediate the 
inflows. In four dimensions, the only states involved in the GS mechanism 
are axionic scalars and their dual 2-forms in the RR sector, belonging 
to linear multiplets~\footnote{Throughout this section, for simplicity 
we will always write anomalous couplings both for scalars and their 
dual 2-forms, despite the well-known fact that there is no local and 
covariant Lagrangian in which a potential and its dual can appear at 
the same time. More precisely, one should write the couplings to the 
dual potential as corrections to the kinetic terms for the field strength 
of the original potential.}.

As anticipated in the introduction, the analysis of the $N=2$ sectors
containing fixed-planes is complicated by the fact that the inflows in the
99 and 55 sectors vanish. Also the $N=4$ untwisted sector present some 
particularities, although it yields a trivially vanishing inflow.
We will therefore analyze separately the $N=1$, $N=2$ and $N=4$ sectors.

\subsection{$N=1$ sectors}

For $N=1$ sectors, we assume that both D9 and D5-branes couple in a 
symmetric way to all the twisted states arising at orbifold fixed-points
contained in their world-volume~\footnote{Clearly, this assumption 
is needed in our factorization procedure, but by a direct computation 
on the disk or crosscap one should be able to find the correct 
combination of fields.}. As we will see, it seems that this quite reasonable 
assumption needs to be relaxed, in order to explain the factorization
in $N=2$ sectors.

\vskip 10pt 
\noindent 
{\bf $\Z_N$ models}
\vskip 5pt 

\noindent
For $\Z_N$ models, the orientifold projection relates the $k$ and 
$(N-k)$-twisted sectors, which together yield always linear multiplets 
\cite{klein}. The distinct twisted sectors are therefore labeled by 
$k=1,...,[(N-1)/2]$. The $N=1$ sectors are those with $k \in S$, where
$$
S = \Big\{k=1,...,[(N-1)/2] \,|\, N_{k} \neq 0 \Big\} \;.
$$
To facilitate the analysis, it is convenient to keep all the 
$N_k$ linear multiplets arising from the $k$ and $(N-k)$-twisted 
sectors as independent states, although in general only 
$N_k^{\prime}\leq N_k$ of them are independent. The physical
$N_k^{\prime}$ propagating states are identified after suitable 
${\bf Z}_N$-projections, as in \cite{ss2,klein}, and are reported 
in the appendix.

In order to figure out which states are responsible for the various 
inflows, recall that the insertion of $\theta^k$ acts as a 
$k$-twist in (\ref{AN}) and as a $2k$-twist in (\ref{MN}), for 
the closed string state exchanged in the transverse channel.
Untwisted closed string exchange can arise only from the $k=0$ part
of (\ref{ZAN}) and the $k=0$ or $k=N/2$ (when present) parts of (\ref{ZMN}).
The $k=0$ part is always trivially vanishing and from the form of $v_3$
(see table 1) it is easy to see that the same is true for the 
$k=N/2$ part. Analogously, no $N/2$-twisted closed string states can 
contribute to the inflow: the relevant contribution is the $k=N/2$ part
of (\ref{ZAN}), which vanishes.
There is therefore a non-vanishing contribution only
from $k$-twisted closed strings, with $k\neq 0,N/2$.
In order to make the corresponding inflow more explicit, it suffice to 
group the $k$ and $N-k$ term of each sum in (\ref{ZAN}) and (\ref{ZMN}). 
The relevant 6-form component of the inflow can then be rewritten as a 
sum over $k\in S$ only, of the following expressions:
\bea
\a\a I_{{\cal A}}^{(k)99}(R,F_9) = \frac{i}{2} N_k\,
Z^9_{(k)}(F_9,R)\,Z^9_{(k)}(F_9,R) \;, \raisebox{16pt}{} \nn \\
\a\a I_{{\cal A}}^{(k)55}(R,F_5) = \frac{i}{2}\,N_k^3\, 
Z^5_{(k)}(F_5,R)\,Z^5_{(k)}(F_5,R) \;, \raisebox{16pt}{} \nn \\
\a\a I_{{\cal A}}^{(k)95}(R,F_5,F_9) = i\, N_k^3\,
Z^9_{(k)}(F_9,R)\,Z^5_{(k)}(F_5,R) \;, \raisebox{16pt}{} \nn \\
\a\a I_{{\cal M}}^{(2k)9}(R,F_9) = i\, N_k\, 
Z^9_{(2k)}(F_9,R)\,Z_{(2k)}(R) \;, \raisebox{20pt}{} \nn \\
\a\a I_{{\cal M}}^{(2k)5}(R,F_5) = i\, N_k^3\,
Z^5_{(2k)}(F_5,R)\,Z_{(2k)}(R) \;, \raisebox{20pt}{} 
\label{fact}
\eea
where
\bea
\a\a Z^5_{(k)}(F_5,R) = \frac {\alpha_k^5}{\sqrt{N}}\,
\sqrt{\Bigg|\frac {s_k^1 s_k^2}{s_k^3}\Bigg|}\,
{\rm ch}(\gamma_k\,\epsilon_k F_5)\,\sqrt{\widehat{A}(R)} \;, \nn  \\
\a\a Z^9_{(k)}(F_9,R) = \frac {\alpha_k^9}{\sqrt{N}}\,
\sqrt{\Bigg|\frac 1 {s_k^1 s_k^2 s_k^3}\Bigg|}\,
{\rm ch}(\gamma_k\,\epsilon_k F_9)\,
\sqrt{\widehat{A}(R)} \;, \raisebox{21pt}{} \nn \\
\a\a Z_{(2k)}(R) = -\frac {4 \, \epsilon_k}{\sqrt{N}}\,
\sqrt{\Bigg|\frac {c_k^1 c_k^2 c_k^3}{s_k^1 s_k^2 s_k^3}\Bigg|}\,
\sqrt{\widehat{L}(R/4)} \;. \raisebox{20pt}{} \label{b2k}
\eea
The coefficients $\epsilon_k$ and $\alpha^\alpha_k$ are suitable signs,
which are required when some of the $s^i_k$ are negative and 
can be chosen as $\epsilon_k = {\rm sign}(s^1_k s^2_k s^3_k)$, 
$\alpha_k^5={\rm sign}(s^3_k)$, $\alpha_k^9 = {\rm sign}(s^1_k s^2_k s^3_k)$.
The corresponding anomalous couplings are
\bea
\a\a S^{(k)}_{D5} = \sqrt{2\pi} \, \sum_{i_k=1}^{N_k^3} 
C_{(k)}^{i_k} \wedge Z_{(k)}^5 \;, \nn \\
\a\a S^{(k)}_{D9} = \sqrt{2\pi} \sum_{i_k=1}^{N_k} 
C_{(k)}^{i_k} \wedge Z_{(k)}^9 \;, \nn \\
\a\a S^{(2k)}_{Fk} = \sqrt{2\pi} \sum_{i_k=1}^{N_k} 
C_{(2k)}^{i_k} \wedge Z_{(2k)} \;, \label{ancN}
\eea
where $C_{(k)}^{i_k} = \chi_{(k)}^{i_k} + \tilde\chi_{(k)}^{i_k}$ is
the sum of the RR axions $\chi_{(k)}^{i_k}$ and their duals 
$\tilde\chi_{(k)}^{i_k}$ at each fixed-point; it is understood that 
one has to keep only the appropriate 6-form component of the integrands. 
The last contribution refers to the k-fixed-points, that indeed do have 
anomalous couplings to the gravitational curvature, in close analogy to 
the six-dimensional case analyzed in \cite{ss2}. The corresponding inflow 
is due to (2k)-twisted axions and $C_{(2k)}=C_{(N-2k)}$ when $2k>N/2$.
Moreover, as ordinary flat orientifold planes, these couplings involve 
the Hirzebruch polynomial ${\hat L(R)}$ \cite{mss}.

Notice that in deriving couplings by factorization there is always 
the arbitrariness of rescaling each $\chi_{(k)}^{i_k}$ and its dual 
$\tilde \chi_{(k)}^{i_k}$ by two opposite factors. Here and in the 
following we write the couplings in the most symmetric way. Furthermore, 
possible couplings that do not give rise to inflow, like couplings only 
to the axion or only to its corresponding tensor field, are obviously
undetectable through this approach.

Recall now that using the condition of non-Abelian anomaly cancellation
(\ref{condN}), the total anomaly (\ref{explN}) simplifies to the 
expression (\ref{simplN}) looking like a pure annulus inflow, 
the M\"obius contribution to the mixed $U(1)$-gravitational anomaly 
being proportional to the corresponding annulus contribution.
This observation and the simple form of the anomaly (\ref{simplN}) 
strongly suggest that the only effect of the fixed-point couplings 
should be to rescale the gravitational couplings of the D-branes. 
One can check that this is indeed what happens. The net effect of the 
fixed-point couplings is simply to rescale by a factor of $3/2$ the 
gravitational part of those of D-branes.
Correspondingly, the total GS term for $N=1$ sectors, obtained by 
adding the three couplings (\ref{ancN}), becomes
\be
S_{GS}^{\Z_N} = \sqrt{\frac{2\pi}{N}} 
\raisebox{2pt}{$\displaystyle{\sum_{k\in S}}$} (N_k)^{1/4} \!\!
\raisebox{2pt}{$\displaystyle{\sum_{\alpha=9,5}}$} \alpha^\alpha_k
\int \left[\chi_{(k)}^\alpha \wedge Y_{(k)}^\alpha
+\tilde\chi_{(k)}^\alpha \wedge X_{(k)}^\alpha \right] \;, 
\label{GSN}
\ee
in terms of the 2 and 4-forms
\bea
\a\a X_{(k)}^\alpha = \frac {i\epsilon_k}{2\pi}
{\rm tr}(\gamma_k F_\alpha) \;, \nn \\
\a\a Y_{(k)}^\alpha = - \frac 1{2(2\pi)^2}
\left({\rm tr}(\gamma_k F_\alpha^2) 
- \frac 1{32} {\rm tr}(\gamma_{k,\alpha})\,{\rm tr} R^2 \right) \;,
\eea
and the normalized combination of axions
\be
\chi_{(k)}^9 =\frac 1{\sqrt{N_k}} \sum_{i_k=1}^{N_k} \chi_{(k)}^{i_k} \;,\;\; 
\chi_{(k)}^5 = \frac 1{\sqrt{N_k^3}} \sum_{i_k=1}^{N_k^3} \chi_{(k)}^{i_k}\;.
\label{defchiN}
\ee
Denoting schematically by $\langle\,\chi \tilde\chi\, \rangle$
the inflow generated by the presence of these fields in the GS term
(\ref{GSN}), one gets $\langle\,\chi_{(k)}^\alpha\,\tilde \chi_{(k)}^\beta 
\,\rangle \sim (N_{k})^{-1/2} \,|C_{k}^{\alpha \beta}|$ according to the 
definitions (\ref{defchiN}), and the inflow induced by (\ref{GSN}) is 
given simply by
\be
I_{GS}^{\Z_N} = \frac{i}{2N} \,\raisebox{2pt}{$\displaystyle{\sum_{k\in S}}$}
\, \raisebox{2pt}{$\displaystyle{\sum_{\alpha,\beta=9,5}}$} \epsilon_k \,
C_{k}^{\alpha \beta}\, X_{(k)}^\alpha \wedge Y_{(k)}^\beta \;,
\ee
reproducing the contribution of $N=1$ sectors to (\ref{simplN}).

\vskip 10pt 
\noindent 
{\bf $\Z_N \times \Z_M$ models}
\vskip 5pt 

\noindent
For $\Z_N \times \Z_M$ orientifolds, the situation is similar.
States which are untwisted with respect to the $\Z_N$ factor 
do not contribute to the inflow. However, one has to consider all the
twists with respect to the $\Z_M$ factor. In this case, the orientifold 
projection relates $(k,l)$ and $(N-k,M-l)$ twists, yielding 
$N_{k,l}^{\prime}$ linear multiplets from each pair of twisted sectors. 
The distinct twisted sectors are then labeled by $k=1,...,[(N-1)/2]$,
$l=0,1,...,M-1$. In general $N_{k,l}^{\prime}\leq N_{k,l}$ but, as before, 
it is convenient to keep all the $N_{k,l}$ fields arising in each twisted 
sector. The $N=1$ sectors come from $(k,l) \in S$, with 
$$
S = \Big\{ k=1,...,[(N-1)/2] \;,\;
l=0,1,...,M-1 \,|\, N_{k,l} \neq 0 \Big\} \;.
$$
Proceeding as in the $\Z_N$ case, one finds perfectly similar results
for the anomalous couplings of D-branes and fixed-points. Again, 
the contribution of the fixed-points can be rewritten in the same form
as the D-brane contributions, by using the condition (\ref{condNM}).
The total GS coupling in the $N=1$ sectors is then 
found to be:
\be
S_{GS}^{\Z_N \times \Z_M} = \sqrt{\frac{2\pi}{NM}} 
\raisebox{2pt}{$\displaystyle{\sum_{(k,l) \in S}}$} (N_{k,l})^{1/4} \!\!
\raisebox{2pt}{$\displaystyle{\sum_{\alpha=9,5}}$} \alpha^\alpha_{k,l} 
\int \left[\chi_{(k,l)}^\alpha \wedge Y_{(k,l)}^\alpha
+\tilde\chi_{(k,l)}^\alpha \wedge X_{(k,l)}^\alpha \right] \;, 
\label{GSNM}
\ee
where
\bea
\a\a X_{(k,l)}^\alpha = \frac {i\epsilon_{k,l}}{2\pi}
{\rm tr}(\gamma_k \delta_l F_\alpha) \;, \nn \\
\a\a Y_{(k,l)}^\alpha = - \frac 1{2(2\pi)^2}\left({\rm tr}
(\gamma_k \delta_l F_\alpha^2) - \frac 1{32} {\rm tr}
(\gamma_{k,\alpha} \delta_{l,\alpha}) \,{\rm tr} R^2 \right) \;,
\eea
and 
\be
\chi_{(k,l)}^9 =\frac 1{\sqrt{N_{k,l}}} \sum_{i_{k,l}=1}^{N_{k,l}} 
\chi_{(k,l)}^{i_{k,l}} \;,\;\; 
\chi_{(k,l)}^5 = \frac 1{\sqrt{N_{k,l}^3}}
\sum_{i_{k,l}=1}^{N_{k,l}^3}  \chi_{(k,l)}^{i_{k,l}}\;,
\label{defchiNM}
\ee
The signs $\epsilon_{k,l}$ and $\alpha^\alpha_{k,l}$ are defined similarly 
to before and one can take the values
$\epsilon_k = {\rm sign}(s^1_{k,l} s^2_{k,l} s^3_{k,l})$, 
$\alpha_{k,l}^5={\rm sign}(s^3_{k,l})$, 
$\alpha_{k,l}^9 = {\rm sign}(s^1_{k,l} s^2_{k,l} s^3_{k,l})$.
Again, one can check that the definitions (\ref{defchiNM}) imply that 
$\langle \, \chi_{(k,l)}^\alpha \, \tilde \chi_{(k,l)}^\beta \, 
\rangle \sim (N_{k,l})^{-1/2} \, |C_{k,l}^{\alpha \beta}|$, and  
the inflow induced by the GS couplings (\ref{GSNM}) is given by 
\be
I_{GS}^{\Z_N \times \Z_M} = \frac{i}{2NM} \,
\raisebox{2pt}{$\displaystyle{\sum_{(k,l) \in S}}$} \,
\raisebox{2pt}{$\displaystyle{\sum_{\alpha,\beta=9,5}}$} \epsilon_{k,l} \,
C_{k,l}^{\alpha \beta}\, X_{(k,l)}^\alpha \wedge Y_{(k,l)}^\beta \;,
\ee
reproducing the contribution of $N=1$ sectors to (\ref{simplNM}).

\subsection{$N=2$ sectors}

$N=2$ sectors, i.e. $k$-twisted sectors containing planes left fixed by 
the orbifold group, can arise when $k v^i = {\rm integer}$ for some $i$, 
in which case $N_k^i = 0$. A subset of these sectors, the $k=N/2$ sectors 
of all the even $\Z_N$ models and the $(k=0,l=3)$ sector of the 
$\Z_3\times\Z_6$ model, do not contribute at all to the inflow, but 
the field theory explanation for this fact is clear and reported in next 
section. Among the remaining sectors, only the $k=2$ sector of the 
${\bf Z}_6^\prime$ model and the ($k=1,l=0,4$) sectors of the 
${\bf Z}_3 \times {\bf Z}_6$ model, give a non-vanishing inflow in the 
95 sector. The above one is reproduced by adding the following couplings:
\bea
\a\a S_{GS}^{\prime\,\Z_6^\prime} = \sqrt{\frac{\pi}{3}}\,3^{1/4} \! 
\raisebox{2pt}{$\displaystyle{\sum_{\alpha=9,5}}$} \alpha^\alpha_{2}
\int \left[\chi_{(2)}^\alpha \wedge Y_{(2)}^\alpha
+\tilde\chi_{(2)}^{\prime\alpha} \wedge X_{(2)}^\alpha \right] \;, 
\label{GSZ6'} \\
\a\a S_{GS}^{\prime\,\Z_3\times\Z_6} = \sqrt{\frac{\pi}{9}} 
\raisebox{2pt}{$\displaystyle{\sum_{l=0,4}}$} 3^{1/4} 
\raisebox{2pt}{$\displaystyle{\sum_{\alpha=9,5}}$} \alpha^\alpha_{1,l}
\int \left[\chi_{(1,l)}^\alpha \wedge Y_{(1,l)}^\alpha
+\tilde\chi_{(1,l)}^{\prime\alpha} \wedge X_{(1,l)}^\alpha \right] \;.
\label{GSZ3Z6}
\eea
The 2 and 4-forms $X$ and $Y$ are the same as before. However,
in order to reproduce a vanishing 99/55 inflow and the correct 95 one, 
two orthogonal combinations of RR fields have to be introduced,
$\chi^\alpha$ and $\chi^{\prime \alpha}$, such that 
$\langle \chi^{9,5} \tilde\chi^{9,5} \rangle =0$ and
$\langle \chi^{9,5} \tilde\chi^{5,9} \rangle =1$.
The crucial difference with the previously analyzed $N=1$ sectors
is that $\tilde\chi^{\prime 9,5}$ is clearly not the dual of $\chi^{9,5}$.

The massless closed string content in these twisted sectors
with non-vanishing inflow is always the same \cite{klein,kr}. 
There are six linear multiplets, but only three of these arise inside 
the world-volume of the D5-branes, the other three being combinations 
of fields living outside the world-volume of the D5-branes.

At this point, one has to make some assumption in order to extract 
a definite answer for the combination of axions $\chi^\alpha$ and 
$\chi^{\prime \alpha}$. A first reasonable assumption is that D5-branes 
couple only to the first three linear multiplets arising within their 
world-volume, whereas D9-branes can couple to all six of them. Another
equally reasonable but less obvious assumption is to require a symmetry 
between all similar fixed-points, as done for the $N=1$ sectors. 
Unfortunately, these two assumptions are together incompatible with the 
inflows computed before, and one of them has necessarily to be relaxed.
In this way, however, even making use of T-duality, which relates the 
D5 and D9-brane couplings to each other, one is left with some free 
parameters which cannot be completely fixed.

We would like to stress that the results of section three nevertheless 
demonstrate anomaly cancellation, and that a net inflow takes place also 
in certain $N=2$ sectors. However, we conclude that the factorization 
approach followed here is not powerful enough to fix unambiguously the 
combinations of fields entering the corresponding anomalous couplings. 
It should be possible to obtain the precise coefficient through a more 
direct computation on the disk and the crosscap.

\subsection{$N=4$ sectors}

A comment is in order also about the untwisted $N=4$ sector.
As shown in section three, the inflow in this sector is trivially vanishing.
However, there are definitely WZ couplings, although they do 
not interfere to give an inflow. This can be understood by looking at this
sector from the $D=10$ Type I point of view and compactify the usual 
anomalous couplings of the D9-branes and the O9-plane on $T^6$, 
and those of the D5-branes and O5-planes (whenever these occur) on a $T^2$. 
Since the curvatures are non-trivial only in four-dimensional non-compact 
space, the only terms surviving this reduction are those obtained by 
integrating the ten-dimensional 6-form on $T^6$, yielding a four-dimensional 
scalar, and the six-dimensional 2-form on $T^2$, yielding another 
four-dimensional scalar. Since no couplings to the two-forms 
dual to these scalars appear, no inflow of anomaly is induced.

\section{Field theory analysis}

One of the consequences of the presence of the GS terms 
(\ref{GSN}), (\ref{GSNM}), (\ref{GSZ6'}) and (\ref{GSZ3Z6})
in the low-energy effective action is the spontaneous 
breakdown of various combinations of $U(1)$ factors 
\cite{dsw,damo,iru1}, involved in the would-be 
$U(1)$-gauge/gravitational anomalies.
The Higgs mechanism is due to the couplings to the 2-forms 
$\tilde\chi^\alpha$ or $\tilde\chi^{\prime\alpha}$,
which modify (after dualization) the kinetic terms for the 
corresponding axions $\chi^\alpha$ or $\chi^{\prime\alpha}$
by a shift involving the $U(1)$ gauge fields. Gauge invariance requires 
then that these fields transform inhomogeneously under the appropriate
gauge transformation, and the scalar is eaten by the gauge field.
The combinations of $U(1)$ gauge fields involved in this mechanism are 
reported in the appendix for each model. 
As usual, care is needed in interpreting string theory results in terms 
of couplings in a low-energy effective action. In our case, particular 
attention is needed because of the well-known chiral multiplet - linear 
multiplet duality \cite{cfv}. We do not discuss this issue here, since
it has has been extensively analyzed in \cite{abd,klein} in the context 
of D=4 $N=1$ IIB orientifolds, and recall simply that the string theory 
results derived in last section can be interpreted most directly in the 
linear multiplet formulation.

The models considered here have $N=1$ supersymmetry. This allows to fix 
the tree-level form of two other CP-even couplings, related
by supersymmetry to the GS couplings above: Fayet-Iliopoulos (FI)
terms~\footnote{In this case, only a tree-level contribution seems to be
present \cite{popp}.} and gauge couplings (GC) depending on the 
NSNS scalars $m^\alpha$, partners of the twisted RR fields. 
As well known, FI terms can trigger both supersymmetry and gauge symmetry 
breaking, and a supersymmetric vacuum with unbroken non-Abelian gauge group 
requires a vanishing FI term. This is particularly important since 
the vacuum expectation values (vev) of the scalars $m^\alpha$ is fixed 
by the FI terms.

Consider first the $N=1$ sectors. The FI terms can be read directly from 
the $\tilde\chi\wedge X$ couplings in the GS terms of last section, and
from the $\chi \wedge Y$ couplings one gets the following GC corrections:
\bea
\a\a  S_{GC}^{\Z_N}= \sqrt{\frac{2\pi}{N}} 
\raisebox{2pt}{$\displaystyle{\sum_{k\in S}}$} (N_k)^{1/4} \!\!
\raisebox{2pt}{$\displaystyle{\sum_{\alpha=9,5}}$} \alpha^\alpha_{k}
\int\! d^4x \,m_{(k)}^\alpha \, 
{\rm tr} (\gamma_k F_{\alpha,\mu\nu} F_\alpha^{\mu\nu}) \;,\label{GCN}\\
\a\a S_{GC}^{\Z_N\times\Z_M}=\sqrt{\frac{2\pi}{NM}} 
\raisebox{2pt}{$\displaystyle{\sum_{(k,l)\in S}}$} 
(N_{k,l})^{1/4} \!\! \raisebox{2pt}{$\displaystyle{\sum_{\alpha=9,5}}$} 
\alpha^\alpha_{k,l} \int\! d^4x \, m_{(k,l)}^\alpha \,
{\rm tr} (\gamma_k\delta_l F_{\alpha,\mu\nu} F_\alpha^{\mu\nu}) \;.
\label{GCNM}
\eea
Importantly, the FI terms and GC involve the same combination of scalars.
For unbroken supersymmetry and non-abelian symmetry, a vanishing FI term 
implies that the vev of the combination of scalars appearing in (\ref{GCN})
and (\ref{GCNM}) vanish as well. Unless one breaks the 
gauge group and/or supersymmetry, there are no gauge coupling corrections.
Apart from an overall coefficient, our results (\ref{GCN})-(\ref{GCNM}) 
reproduce those of \cite{abd}, extending them to other models and 
gauge fields in the 55 sector. 

As expected, $N=2$ sectors are more subtle. For $Z_2$-twisted
$N=2$ sectors, which have fixed-planes along the world-volume of 
the D5-branes, the inflow vanishes. This fact alone is not sufficient 
to rule out corrections to gauge couplings, but supersymmetry implies 
that couplings like (\ref{GCN})-(\ref{GCNM}) are nonetheless forbidden. 
Indeed, for these sectors the compact space is effectively 
$R^4\times T^2\times T^4/G$, with $N=2$ supersymmetry and the D5-branes 
wrapped on $T^2$, and one can therefore treat them as $\Z_2$-twisted 
sectors of dimensionally reduced $D=6$ $N=1$ orientifolds. 
In these sectors the only RR states exchanged are scalars, belonging
to $D=6$ hypermultiplets \cite{ss2}. Since these cannot 
couple to gauge kinetic terms in $D=6$ \cite{dW}, these couplings are 
forbidden also in the $D=4$ $N=1$ model.
The situation is different for $N=2$ sectors with fixed-planes outside 
the D5-branes world-volume. In these cases, the inflow is either vanishing 
($k=3$ in $\Z_{12}$ and $(k=0,l\neq 0,3)$ in $\Z_3\times\Z_6$)
or restricted to the 95 sector ($k=2$ in $\Z_6^\prime$ 
and $(k=1,l=0,4)$ in $\Z_3\times\Z_6$).
In the first case, a dependence of the gauge couplings on the 
corresponding scalars is allowed if no FI term is generated at all.
In the second case, the combination of scalars appearing in the 
gauge couplings and the FI terms have to be orthogonal, and again 
a gauge coupling dependence is allowed. Summarizing, in the orbifold 
limit where all these scalars have zero vev, no gauge coupling correction 
is present in any case, but there is the interesting possibility of 
orbifold deformations leading to a non-vanishing tree-level correction 
to the gauge couplings, even for unbroken gauge symmetry. 
A further study is definitely needed to get a better understanding of 
these sectors. 

Analogously to the $F\wedge F$ terms, supersymmetry, or better 
supergravity, also fixes the form of the various CP-even terms related 
to the $R\wedge R$ couplings. Being higher order in derivatives, we 
will not discuss the explicit form of these terms.


\acknowledgments{We wish to thank G. Lopes Cardoso for continous and fruitful
discussions, L. Ibanez and M. Klein for important correspondence, and 
J.-P. Derendinger and C. Bachas for interesting comments and suggestions. 
This work has been supported by the EEC under TMR contract ERBFMRX-CT96-0045 
and by the Fundamenteel Onderzoek der Materie (FOM).}


\appendix

\section{Appendix}

We report here the explicit combination of the various
would-be anomalous $U(1)$ factors that become massive for each model.
In order to find their right combination, it is important
to write the GS couplings of section four in terms of physical
fields only. All the combinations are symmetric in the $9,5$ exchange, 
as expected by T-duality.

\vskip 10pt 
\noindent 
{\bf $\Z_3$}
\vskip 5pt

\noindent
All the fields appearing in (\ref{GSN}) are physical and 
the unique $U(1)$ present in the model gets a mass.

\vskip 10pt 
\noindent 
{\bf $\Z_7$}
\vskip 5pt

\noindent
Again, all the 7 fields appearing in each of the three
sectors in (\ref{GSN}) are physical. An independent combination of $U(1)$
factors is present in each sector. Correspondingly, all of the three $U(1)$'s
present in the model get a mass.

\vskip 10pt 
\noindent 
{\bf $\Z_6$}
\vskip 5pt

\noindent
The 3 fields in the $k=1$ sector are all physical. In the
$k=2$ sector, $\chi_{(2)}^{1,2,3}$ are again all physical,
whereas the other 24 fields give rise to only 12 propagating
combinations given by $1/\sqrt{2}(\chi_{(2)}^i+\chi_{(2)}^{i+12})$,
$i=4,...,15$. The three massive $U(1)$ fields are 
\bea
\a\a A_{(1)}= {\rm tr} (\gamma_1 A_9) + {\rm tr} (\gamma_1 A_5) \;, \nn \\
\a\a A_{(2,3)}= {\rm tr} (\gamma_2 A_{9,5}) + \frac 13 {\rm tr} 
(\gamma_2 A_{5,9}) \;. \nn
\eea

\vskip 10pt 
\noindent 
{\bf $\Z_6^\prime$}
\vskip 5pt

\noindent
All the fields are physical. The four massive $U(1)$'s are
\bea
\a\a A_{(1,2)}= {\rm tr} (\gamma_1 A_{5,9}) - \frac 12 {\rm tr} 
(\gamma_1 A_{9,5}) \;, \nn \\
\a\a A_{(3,4)}= {\rm tr} (\gamma_2 A_{9,5})+ c \, {\rm tr} (\gamma_2 A_{5,9})  
\;, \raisebox{12pt}{} \nn 
\eea
where $c$ is an arbitrary coefficient.
The form of the gauge fields $A_{(3,4)}$ is dictated by T-duality only,
since they get a mass by higgsing axions in $N=2$ sectors, where our 
factorization procedure is ambigous.

\vskip 10pt 
\noindent 
{\bf $\Z_{12}$}
\vskip 5pt

\noindent
Fields are identified under a $\Z_N$ projection only in the
$k=4$ sector. In this case, $\chi_{(4)}^{1,2,3}$ are physical, whereas 
the other 24 fields give rise to 6 propagating combinations given by 
$1/2(\chi_{(4)}^i+\chi_{(4)}^{i+6}+\chi_{(4)}^{i+12}+\chi_{(4)}^{i+18})$,
i=4,...,9. The five massive gauge fields are
\bea
\a\a A_{(1,2)} = {\rm tr} (\gamma_{1,2} A_{9}) - 
{\rm tr} (\gamma_{1,2} A_{5}) \;, \nn \\
\a\a A_{(3,4)} = {\rm tr} (\gamma_4 A_{9,5}) + \frac 13  
{\rm tr} (\gamma_4 A_{5,9}) \;, \nn \\
\a\a A_{(5)} = {\rm tr} (\gamma_5 A_{9}) + {\rm tr} (\gamma_5 A_{5}) 
\raisebox{12pt}{} \;. \nn
\eea

\vskip 10pt 
\noindent 
{\bf $\Z_3 \times\Z_3$}
\vskip 5pt

\noindent
All the 27 fields are physical and a single $U(1)$ gets a mass:
$$
A= {\rm tr} (\gamma_1\delta_2 A_9) \;.
$$

\vskip 10pt 
\noindent 
{\bf $\Z_3 \times\Z_6$}
\vskip 5pt

\noindent
Fields are identified in the $(k,l)=(1,1),(1,2),(1,3)$ sectors.
In all these sectors, the first three states, {\em i.e.} those coupling
also to D5-branes, are physical. The remaining physical combinations
are $1/\sqrt{3}(\chi_{(1,1)}^{i}+\chi_{(1,1)}^{i+3}+\chi_{(1,1)}^{i+6})$,
$i=4,5,6,\;$ $1/\sqrt{2} (\chi_{(1,2)}^{i}+\chi_{(1,2)}^{i+12})$, 
$i=4,...,12$, and $1/\sqrt{3} (\chi_{(1,3)}^{i}+
\chi_{(1,3)}^{i+3}+\chi_{(1,3)}^{i+6})$, $i=4,5,6$.
The eleven massive $U(1)$'s are 
\bea
\a\a A_{(1,2)}= {\rm tr} (\gamma_1 A_{9,5})+ d \, {\rm tr} (\gamma_1 A_{5,9}) \;, \nn \\
\a\a A_{(3,4)} = {\rm tr} (\gamma_1\delta_1 A_{9,5}) - \frac 12
{\rm tr} (\gamma_1\delta_1 A_{5,9}) \;, \raisebox{19pt}{} \nn \\
\a\a A_{(5,6)} = {\rm tr} (\gamma_1\delta_2 A_{9,5}) - \frac 13  
{\rm tr} (\gamma_1\delta_2 A_{5,9}) \;, \nn \\
\a\a A_{(7,8)} = {\rm tr} (\gamma_1\delta_3 A_{9,5}) - \frac 12  
{\rm tr} (\gamma_1\delta_3 A_{5,9}) \;, \nn \\
\a\a A_{(9,10)}= {\rm tr} (\gamma_1\delta_4 A_{9,5}) + e \,
{\rm tr} (\gamma_1\delta_4 A_{9,5}) \;, 
\raisebox{15pt}{} \nn \\ 
\a\a A_{(11)} = {\rm tr} (\gamma_1\delta_5 A_{9}) + 
{\rm tr} (\gamma_1\delta_5 A_{5}) \;. \raisebox{19pt}{} \nn
\eea
Again, the arbitrary coefficients $d$ and $e$ reflect the ambiguity
in the factorization procedure for $N=2$ sectors. For this reason, 
we have not explicitly checked whether the gauge fields above are 
always independent or not.


\end{document}